\documentclass[12pt]{article}

\font\small=cmr10
\usepackage{color}

\usepackage{amsmath}
\usepackage{amsfonts,amssymb}
\newcommand{\h}{{\mathsf{h}}}
\newcommand{\mi}{{\mathrm{i}}}

\newcommand{\set}[1]{\left\{ #1\right\}}
  
\newcommand{\tr}[1]{{\rm tr }\left(#1\right)}  
\newcommand{\Tr}[1]{{\rm Tr }\left(#1\right)}

\newcommand{\bo}[1]{\mathcal{B}(#1)}

\newcommand\unit{\hbox{\rm 1\kern-2.8truept l}}
\newcommand{\Ll}{{\mathcal{L}}}
\newcommand{\Tt}{{\mathcal{T}}}

\newcommand{\avanti}{\overrightarrow{\Omega}}
\newcommand{\indietro}{\overleftarrow{\Omega}}

\newcommand{\Vava}[1]{\overrightarrow{V}_{#1}}
\newcommand{\Vind}[1]{\overleftarrow{V}_{#1}}

\newcommand{\phiava}{{\overrightarrow{\Phi}_*}}
\newcommand{\phiind}{{\overleftarrow{\Phi}_*}}

\newcommand{\ketbra}[1]{\left\vert #1\right\rangle\left\langle #1\right\vert}

\newcommand{\eprod}[1]{{\mathbf{ep}(#1)}}

\newtheorem{definition}{Definition}
\newtheorem{theorem} {Theorem}

\newtheorem{lemma} {Lemma}
\newtheorem{corollary}{Corollary}

\begin{document}


\title{\bf Entropy production and detailed balance 
for a class of quantum Markov semigroups}
%
%
%
%
\date{}
\maketitle
\author{}
\centerline{{F. Fagnola}} 
\centerline{\small{Dipartimento di Matematica, Politecnico di Milano}} 
\centerline{\small{Piazza Leonardo da Vinci 32, I-20133 Milano (Italy)}}
\centerline{\small{franco.fagnola@polimi.it}}
\bigskip
\centerline{{R. Rebolledo}} 
\centerline{\small{Centro de An\'alisis Estoc\'astico y Aplicaciones }} 
\centerline{\small{ Facultad de Ingenier{\'\i}a 
y Facultad de Matem\'aticas }} 
\centerline{\small{Pontificia Universidad Cat\'olica de Chile}} 
\centerline{\small{Casilla 306, Santiago 22, Chile.}}
\centerline{\small{rrebolle@uc.cl}}

\begin{abstract}
We give an explicit entropy production formula for a class of quantum 
Markov semigroups, arising in the weak coupling limit of a system  coupled with reservoirs, whose generators $\Ll$ are sums of other 
generators $\Ll_\omega$ associated with positive Bohr frequencies 
$\omega$ of the system. 
As a consequence, we show that any such semigroup satisfies the 
quantum detailed balance condition with respect to an invariant  
state if and only if all semigroups generated by each $\Ll_\omega$ 
so do with respect to the same invariant state.
\end{abstract}

\section{Introduction}


Steady states of an open quantum system are  considered 
equilibrium or non equilibrium states according to whether 
or not they satisfy a quantum detailed balance condition (see 
\cite{Agarwal,Alicki-Lendi,FFVU07,FFVU08,FFVU,FFVU12,KFGV,MaSt,Seif} 
and the references therein). 
Concepts of entropy production have been proposed in several papers 
(\cite{BolQue,Breuer,FFRR-QP29,FFRR-ep,JP-2001,MRM,Seif} is a short list far from being complete) 
as an index of deviation from detailed balance
(see \cite{Qian2003} also for classical Markov processes).

In \cite{FFRR-QP29,FFRR-ep} we introduced a definition 
of entropy production rate for faithful normal invariant states of 
quantum Markov semigroups, inspired by the one brought 
into play for classical Markov processes, by considering the 
derivative of relative entropy of the one-step 
forward and backward two-point states at time $t=0$.  
Moreover, we proved an explicit formula for the entropy production 
of a quantum Markov semigroup in terms of the completely positive part of the generator (Theorem \ref{th:ep-formula}  here).
This formula shows that non zero entropy production is closely 
related with the violation of quantum detailed balance conditions 
and singles out states with finite entropy production as a rich class 
of simple non equilibrium invariant states.

In this paper we compute the entropy production for a class of 
quantum Markov semigroups, a faithful invariant state $\rho$, arising 
in the weak coupling limit of a system  coupled with reservoirs, 
whose generators $\Ll$ are sums of other generators $\Ll_\omega$ associated with positive Bohr frequencies $\omega$ 
of the system (see \cite{AcLuVo,Dav,DeFr}).

Our main result is the explicit formula (\ref{eq:ep-formula}) for 
the entropy production rate in terms of second order moments 
of Kraus operators in the GKSL representation of the generator.
This formula shows that the entropy production of a semigroup
in this class is the \emph{sum} of non-negative entropy 
productions of all semigroups generated by each $\Ll_\omega$. 
As a consequence (Theorem \ref{th:GDB-trivial}) the semigroup 
generated by $\Ll$ satisfies the quantum detailed balance 
condition if and only if so does each semigroup generated 
an $\Ll_\omega$.

The plan of the paper is as follows. In Section \ref{sect:QMSclass} 
we introduce the class of quantum Markov semigroups we are 
dealing with. In Section \ref{sect:QDB-ep} we recall various 
notions of quantum detailed balance. Our new formula for 
the entropy production is proved in Section \ref{sect:ep} and, 
finally, in Section \ref{sect:global-local} we essentially show that 
equilibrium states for the semigroup generated by $\Ll$ are 
equilibrium state for all semigroups  generated by each $\Ll_\omega$.

\section{QMS of stochastic limit type}
\label{sect:QMSclass}

We will be concerned with the class of quantum Markov semigroups 
(QMS) we describe below under some restrictive assumptions in 
order to avoid domain problems and similar technicalities. 
This class arises in the weak coupling 
limit as well as in the stochastic limit of a Hamiltonian system $S$ 
interacting with a reservoir (see \cite{AcLuVo,Dav,DeFr} and the 
references therein).

Let $\h$ be a fixed $d$-dimensional ($d<\infty$) Hilbert space  
and let $H_S$ be a self-adjoint operator on $\h$ with spectral 
decomposition
\[
H_S=\sum_n\varepsilon_n P_{\varepsilon_n}
\]
where $\varepsilon_n\not=\varepsilon_m$ for $m\not=n$ and 
$P_{\varepsilon_n}$ is the orthogonal projection onto the nullspace 
of $H_S - \varepsilon_n \unit_\h$ (here $\unit_h$ denotes the 
identity operator on $\h$). We denote as $\mathcal{B}(\h)$ 
the algebra of all bounded operators on $\h$. 
We call \emph{Bohr frequencies} the differences
\[
\omega=\varepsilon_n-\varepsilon_m
\qquad \text{with \ $\varepsilon_n>\varepsilon_m$}
\]

Choose an operator $V$ on $\h$ and define
\begin{equation}\label{eq:V-omega}
V_\omega=\sum_{\varepsilon_n-\varepsilon_m=\omega}
P_{\varepsilon_m}VP_{\varepsilon_n}.
\end{equation}
Moreover, let $H_\omega$ be a self-adjoint operator on $\h$ 
commuting with $H_S$. For all Bohr frequency $\omega$ let 
$\Ll_\omega$ be the GKSL (Gorini-Kossakowski-Sudarshan-Lindblad) generator of a QMS on $\mathcal{B}(\h)$ 
\begin{eqnarray}\label{eq:slt-local}
\Ll_\omega(x)&=&\mi [H_\omega,x]-\frac{\gamma^-_\omega}{2}\left(V_\omega^*V_\omega x-2V^*_\omega xV_\omega+xV_\omega^*V_\omega \right)\nonumber\\
&-&\frac{\gamma^+_\omega}{2}\left(V_\omega V_\omega^* x-2V_\omega xV_\omega^*+xV_\omega V_\omega^* \right).
\end{eqnarray}
where $\gamma_\omega^-,\gamma_\omega^+>0$.
QMSs in our class are generated by the linear map $\Ll$  
\begin{equation}\label{eq:slt-global}
\Ll=\sum_\omega \Ll_\omega.
\end{equation}
Note that, defining
\begin{equation}\label{eq:G-global}
G_\omega=-\frac{1}{2}\left (\gamma^-_\omega V^*_\omega V_\omega+\gamma^+_\omega V_\omega V^*_\omega\right)-\mi H_\omega
\end{equation}
we can write the generator $\Ll_\omega$ simply as 
\begin{equation*}
\Ll_\omega(x) = G_\omega^* x 
+ \gamma^-_\omega V^*_\omega x V_\omega
+\gamma^+_\omega V_\omega x V^*_\omega
+ x G_\omega.
\end{equation*}

Since the Hilbert space $\h$ is finite dimensional the QMS generated 
by $\Ll_\omega$ admits an invariant state $\rho$. Moreover, it is 
well-known (see e.g. \cite{AcLuVo}) that there exists an invariant 
state whose density matrix $\rho$ commutes with the system 
Hamiltonian $H$ so that it can be written as 
\[
\rho = \sum_{1\le j\le d} |e_j\rangle\langle e_j| 
\]
where $\rho_j\ge 0$, the above sum is finite, 
$\sum_{1\le j\le d}\rho_j=1$,  $(e_j)_{1\le j\le d}$ is an
orthonormal basis of $\h$ and each $e_j$ belongs to an 
eigenspace $P_n$ of $H_S$. We shall also assume that $\rho$
is faithful (if not we can reduce the semigroup by its recurrent 
projection \cite{FFRR-LNM1882}).  

The generators of these QMSs turn out to admit a special 
GKSL representantion (\cite{Partha} Theorem 30.16)
\begin{equation}\label{eq:GKSL}
\Ll(x)=\mi[H,x] - \frac{1}{2}
\sum_{\ell =1}^{2b}
\left(L^*_\ell L_\ell x-2L^*_\ell xL_\ell + xL^*_\ell L_\ell\right)
\end{equation}
where $b$ is the number of Bohr frequencies, such that 
$\hbox{\rm tr}(\rho L_\ell)=0$ for all $1\le \ell \le b$ 
and operators $(L_\ell)_{\ell\ge 1}$ are linearly independent 
in $\mathcal{B}(\h)$. Indeed, it suffices to associate with each 
Bohr frequency $\omega$ a pair of operators
\begin{equation}\label{eq:GKSL-L}
L_{2\ell}=(\gamma^-_\omega)^{1/2}V_\omega\qquad L_{2\ell-1}=(\gamma^+_\omega)^{1/2}V_\omega^*,
\end{equation}
where the indexes  run over a finite set, 
and define $H=\sum_\omega H_\omega$.

\section{Quantum detailed balance and entropy production}
\label{sect:QDB-ep}

A number of conditions called \emph{quantum detailed balance} 
(QDB) have been proposed in the literature to distinguish,  among invariant states, those enjoying reversibility properties. 

The first one, to the best of our knowledge, appeared in  the work 
of Agarwal \cite{Agarwal} in 1973 (see also Majewski \cite{Maje}) 
and involves a reversing operation  $\Theta:\mathcal{B}(\h)
\to\mathcal{B}(\h)$, namely a linear $*$-map (\,$\Theta(x^*)= \Theta(x)^*$ for all $x\in\mathcal{B}(\h)$), that is also an  antihomomorphism (\,$\Theta(xy)=\Theta(y)\Theta(x)$\,) 
and satisfies $\Theta^2=I$, where $I$ denotes the identity 
map on $\mathcal{B}(\h)$. A QMS $\mathcal{T}$ satisfies the 
Agarwal-Majewski QDB condition with respect to a faihtful 
normal invariant state $\rho$ if  $\tr{\rho x\Tt_t(y) } = 
\tr{\rho  \,\Theta(y)\Tt_t(\Theta(x))}$,
for all $x,y\in\mathcal{B}(h)$. If the state $\rho$  is 
invariant under the reversing operation, i.e. 
$\tr{\rho\Theta(x)}=\tr{\rho x}$ for all $x\in\mathcal{B}(\h)$, 
as we shall assume throughout the paper,
this condition can be written in the equivalent form 
$\tr{\rho x\Tt_t(y)}= 
\tr{\rho \left((\Theta\circ\Tt_t\circ\Theta)(x)\right) y}$ 
for all $x,y\in\mathcal{B}(h)$. 
Therefore the Agarwal-Majewski QDB condition means that 
maps $\Tt_t$  admit dual maps coinciding with 
$\Theta\circ\Tt_t\circ\Theta$ for all $t\ge 0$; in 
particular dual maps must be positive since 
$\Theta$ is obviously positivity preserving. The map 
$\Theta$ often appears in the physical literature 
as a parity map; a self-adjoint $x$ is an even (resp. odd)
observable if $\Theta(x)=x$ (resp.  $\Theta(x)=-x$). 

In our framework, since $H_S$ is the energy of the system, which 
is a typical even observable,  a reasonable map $\Theta$ is the transpose $\Theta(a)=a^\intercal$ with respect to an orthonormal 
basis $(e_j)_{1\le j\le d}$ of $\h$ diagonalizing $H_S$ as 
in \cite{DuSn}. Interested readers can consult 
\cite{FFVU08,FFVU,Maje} in  more general situations.

The best known QDB notion, however, is due to Alicki \cite{Alicki-Lendi} 
and Kossakowski, Frigerio, Gorini, Verri \cite{KFGV}. According to these 
authors, a QMS with generator $\Ll$ as in \eqref{eq:GKSL}, with 
invariant state $\rho$ whose density commutes with $H$,  satisfies 
the quantum detailed balance condition if 
$\tr{\rho\, x\Ll(y)}=\tr{\rho\,\widetilde{\Ll}(x)y}$
where $\widetilde{\Ll}=\Ll - 2\mi[H,\cdot]$.
As a consequence, the QMS $\widetilde{\Tt}$ on $\mathcal{B}(\h)$ 
generated by $\widetilde{\Ll}$ satisfies $\tr{\rho\, x\Tt_t(y)}
=\tr{\rho\,\widetilde{\Tt}_t(x)y}$ for all $t\ge 0$.

Both the above QDB conditions depend in a crucial way from the 
bilinear form $(x,y)\to \tr{\rho xy}$.  In particular, if they hold true,   
all positive maps $\Tt_t$ admit \emph{positive} dual maps; 
as a consequence, all the maps $\Tt_t$ must commute 
with the modular group $(\sigma^\rho_t)_{t\in\mathbb{R}}$, 
given by $\sigma^\rho_t(x) =\rho^{\mi t} x \rho^{-\mi t}$,  
associated with the state $\rho$ (see \cite{KFGV} Prop. 2.1, 
\cite{MaSt} Prop. 5 and also \cite{Cipriani}) as well as the 
generator $\Ll$. 
This algebraic restriction is unnecessary if we consider the
bilinear form $(x,y)\to \omega\left(\sigma_{\mi/2}(x)y \right)$ 
for defining dual QMSs.

QDB conditions arising when we consider this bilinear form are 
called \emph{standard} (see e.g. \cite{DeFr}, \cite{FFVU});  
we could not find them in the literature, but  it seems that they 
belong to the folklore of the subject. In particular, they were 
considered by R. Alicki and A. Majewski (private communication). 

\begin{definition}\label{def:SQDB}
Let $\Tt$ be a QMS with a dual   $\Tt^\prime$  defined by 
$\omega\left(\sigma_{\mi/2}(x)\Tt_t(y) \right) =\omega
\left(\sigma_{\mi/2}\left(\Tt_t^\prime(x)\right)y \right)$
for all  $x,y\in\mathcal{B}(\h)$, $t\ge 0$. The semigroup $\Tt$ 
satisfies:
\begin{enumerate}
\item the standard quantum detailed balance 
condition with respect to the reversing operation $\Theta$ 
(SQBD-$\Theta$) if $\Tt_t^\prime= 
\Theta\circ\Tt_t\circ\Theta$ for all $t\ge 0$,
\item the standard quantum detailed balance condition (SQDB) 
if the difference of generators $\Ll -\Ll^\prime$ of $\Tt$ and 
$\Tt^\prime$ is a  derivation.
\end{enumerate}
\end{definition}

It is worth noticing here that the above \emph{standard} QDB 
conditions coincide with the Agarwal-Majewski and Alicki-Gorini-Kossakowski-Frigerio-Verri respectively when the  
QMS $\Tt$ commutes with the modular group 
$(\sigma_t)_{t\in\mathbb{R}}$ associated $\omega$ (see \cite{FFVU07,FFVU}).

In the framework of the present paper all states are normal will and 
be identified with their densities.
In particular, $\omega(x)=\tr{\rho\, x}$, $\sigma_t(x)= 
\rho^{\mi t} x \rho^{-\mi t}$ and  
$\omega\left(\sigma_{\mi/2}(x)y \right)
=\tr{\rho^{1/2} x \rho^{1/2}y}$. 

In \cite{FFVU} (Theorems 5, 8 and Remark 4) we proved the 
following characterisations of QMS satisfying a standard QDB 
condition we recall here in the present framework.

\begin{theorem}\label{th:SQDB} 
A QMS $\Tt$ satisfies the SQDB if and only if for any special 
GKSL representation of the generator $\Ll$ by means of 
operators $G,L_\ell$ there exists a unitary 
$(u_{m\ell})_{1\le m, \ell\le 2b}$ 
on $\mathsf{k}$ which is also symmetric (i.e. $u_{\ell m}=
u_{\ell m}$ for all $m,\ell$) such that, for all $\ell\ge 1$,
\begin{equation}\label{sqdb-cond}
\rho^{1/2}L^*_\ell=\sum_{1\le m\le 2b} u_{\ell m}L_m\rho^{1/2}.
\end{equation}
\end{theorem}

\begin{theorem}\label{th:SQDB-TR}
A QMS $\Tt$ satisfies the {SQBD-$\Theta$} condition if and only 
if for any special GKSL representation of $\Ll$ by means   
of operators $G, L_\ell$, there exists a self-adjoint unitary 
$(u_{\ell m})_{1\le m, \ell\le 2b}$  such that:
\begin{enumerate}
\item \label{sdb-theta-1} $\rho^{1/2}G^\intercal =G\rho^{1/2} $, 
\item \label{sdb-theta-2} $\rho^{1/2} L_\ell^\intercal 
      = \sum_{1\le m\le 2b}  u_{\ell m} {L_m}\rho^{1/2}$ 
     for all $1\le \ell \le 2b$.
\end{enumerate}
\end{theorem}

 
The SQBD-$\Theta$ condition is more restrictive than 
the SQDB condition because it involves also the identity 
$\rho^{1/2}G^\intercal=G\rho^{1/2} $ (see 
Example 7.3 in \cite{FFRR-ep}).
However,  this does not happen if $G^\intercal = G$ 
and $\rho$ commutes with $G$. This is a reasonable physical 
assumption satisfied by many QMSs as, for instance,  those
of stochastic limit type we are considering in this paper. 
Conditions obtained including the reversing map $\Theta$ 
seem more suitable for studying quantum detailed balance (\cite{FFVU,JP-2013}).

\section{A formula for entropy production}\label{sect:ep}

We begin this section by recalling our notion of entropy production 
\cite{FFRR-ep}.  Since it provides an index describing deviation from 
detailed balance, it was introduced in \cite{FFRR-QP29, FFRR-ep} 
through the forward and backward two-point states on 
$\mathcal{B}(\h)\otimes\mathcal{B}(\h)$ 
\begin{eqnarray*}
\avanti_t\left( x \otimes y\right) 
& = & \tr{ \rho^{1/2} x^\intercal \rho^{1/2} \Tt_t(y)}\\
\indietro_t\left( x \otimes y\right) & = & 
\tr{ \rho^{1/2} \Tt_t(x^\intercal)^\intercal \rho^{1/2} \Tt_t(y)},
\end{eqnarray*}
which clearly coincide if and only if $\Tt$ satisfies the 
SQDB-$\Theta$ condition, and their relative entropy 
$S(\avanti_t,\indietro_t)$ as 
\begin{equation}\label{eq:ep-def}
\eprod{\Tt,\rho} =
\limsup_{t\to 0^+}\frac{S(\avanti_t,\indietro_t)}{t}
\end{equation}



Moreover, in \cite{FFRR-ep} Theorem 5, we proved an explicit formula 
based on the Kraus operators $L_\ell$ in a GKSL decomposition of 
the generator $\Ll$. Let $\phiava$ and $\phiind$ be the linear maps 
on trace class operators on $\h\otimes\h$
\begin{eqnarray}
\phiava(X)&=& \sum_{\omega}\left(\gamma^-_\omega\left(\unit\otimes V_\omega\right) 
X \left(\unit\otimes V_\omega^*\right)+\gamma^+_\omega\left(\unit\otimes V_\omega^*\right) 
X \left(\unit\otimes V_\omega\right)\right)\label{eq:ava}\\
\phiind(X) &= &\sum_{\omega}\left(\gamma^-_\omega\left(V_\omega\otimes\unit \right) 
X \left(V_\omega^*\otimes\unit\right)
+\gamma^+_\omega\left(V_\omega^*\otimes\unit\right) 
X \left( V_\omega\otimes\unit\right)\right)\label{eq:ind}
\end{eqnarray}
Let $\theta$ be the antilinear conjugation in a basis 
$(e_j)_{1\le j\le d}$ diagonalizing $\rho$ and let $D$ be the 
entangled state on $\mathcal{B}(\h)\otimes\mathcal{B}(\h)$ introduced in \cite{FFRR-ep} as
\begin{equation}\label{eq:density_Omega0}
D=\ketbra{r}, \qquad 
r= \sum_{j}\rho_j^{1/2} \theta e_j\otimes e_j.
\end{equation}
It is not hard to check as in \cite{FFRR-ep} that $D$ is the 
density of the state $\avanti_0=\indietro_0$.

\begin{theorem}\label{th:ep-formula} 
Let $\Tt$ be QMS on $\bo{\h}$ and $\rho$ a faithful invariant state. 
Assume:
\begin{enumerate}
\item $\rho^{1/2}G^\intercal=G\rho^{1/2}$,
\item the linear spans of $\set{L_\ell \rho^{1/2}\,\mid\, \ell\ge 1}$ 
and $\set{ \rho^{1/2} L_\ell^\intercal \,\mid\, \ell\ge 1}$ coincide. 
\end{enumerate}
Then the ranges of  $\phiava(D)$  and $\phiind(D)$ coincide and 
the entropy production is
\begin{eqnarray}\label{eq:ep-formula}
\eprod{\Tt,\rho}& =& \frac{1}{2}
\Tr{\left(\phiava(D)-\phiind(D)\right)
\left(\log\left(\phiava(D)\right) 
- \log\left(\phiind(D)\right)
\right)}\nonumber\\
& =& 
\Tr{\phiava(D)
\left(\log\left(\phiava(D)\right) 
- \log\left(\phiind(D)\right)
\right)}
\end{eqnarray}
\end{theorem}

In order to compute explicitly the entropy production for QMS in 
the class described in Section \ref{sect:QMSclass} we begin by 
establishing a preliminary Lemma. We denote by $\langle \cdot, 
\cdot \rangle$ the scalar product in $\h\otimes \h$.

\begin{lemma}\label{lem:XY}
Let $X$ and $Y$ be bounded operators on $\h$, then
\begin{eqnarray}
\langle (Y\otimes\unit)r,(\unit\otimes X)r\rangle &=&\tr{(\rho^{1/2}\theta Y^*\theta )^*X\rho^{1/2}}\\
\langle (\unit\otimes Y)r,(\unit\otimes X)r\rangle &=&\tr{\rho\, Y^*X}.
\end{eqnarray}
\end{lemma}
\noindent{\it Proof.} Both formulas follow from straightforward 
computations
\begin{eqnarray*}
\langle (Y\otimes\unit)r,(\unit\otimes X)r\rangle &=&\sum_{j,k}(\rho_j\rho_k)^{1/2}\langle Y\theta e_j,\theta e_k\rangle \langle e_j,Xe_k\rangle\\
&=&\sum_{j,k}\langle e_k,\theta Y\theta\rho^{1/2}e_j\rangle \langle e_j,X\rho^{1/2}e_k\rangle\\
&=&\sum_{j,k}\langle \rho^{1/2}\theta Y^*\theta e_k,e_j\rangle \langle e_j,X\rho^{1/2}e_k\rangle\\
&=&\sum_{k}\langle \rho^{1/2}\theta Y^*\theta e_k,X\rho^{1/2}e_k\rangle\\
&=&\tr{(\rho^{1/2}\theta Y^*\theta )^*X\rho^{1/2}}
\end{eqnarray*}
\begin{eqnarray*}
\langle (\unit\otimes Y)r,(\unit\otimes X)r\rangle &=&\sum_{j,k}(\rho_j\rho_k)^{1/2}\langle\theta e_j,\theta e_k\rangle \langle Ye_j,Xe_k\rangle\\
&=&\sum_j\langle  Y \rho^{1/2}e_j, X\rho^{1/2}e_j\rangle\\
&=&\tr{\rho\, Y^*X}
\end{eqnarray*}
\hfill $\square$

Replacing the operators $X,Y$ in Lemma \ref{lem:XY} by 
operators $V_\omega$ and keeping into account that 
$\theta V_\omega^* \theta=V_\omega^*$ if $V_\omega$ 
is a real matrix, we have the following

\begin{corollary}\label{cor:scalar-prod} If the operators  $V_\omega$, 
defined by \eqref{eq:V-omega}, are represented by real matrices we 
have 
\begin{eqnarray*}
\langle (V_{\omega^\prime}\otimes\unit)r, (\unit\otimes V_{\omega})r\rangle&=&\delta_{\omega, \omega^\prime}\tr{\rho^{1/2}V_{\omega}\rho^{1/2}V_{\omega}}\\
\langle (\unit\otimes V_{\omega^\prime})r,(\unit\otimes V_{\omega})r\rangle &=&\delta_{\omega, \omega^\prime} \tr{\rho V_{\omega}^*V_\omega}
\end{eqnarray*}
where $\delta_{\omega,\omega^\prime}$ is the Dirac delta.
\end{corollary}

We are now in a position to prove our entropy production formula 

\begin{theorem}\label{th:ep-QMS-SLT}
Assume that $V_\omega$ and $H_\omega$ are real matrices  for 
all Bohr frequency $\omega$,  then the entropy production is
\begin{eqnarray}\label{eq:ep-formula}
\eprod{\Tt,\rho}& =& 
\sum_\omega\left( \gamma^-_\omega\, \tr{\rho V^*_\omega V_\omega}\log\left( \frac{\gamma^-_\omega \,
\tr{\rho V^*_\omega V_\omega}^2}{\gamma^+_\omega\,\tr{\rho^{1/2}V^*_\omega\rho^{1/2}V_\omega}^2}\right)  \right.\nonumber\\
&+&\left. \gamma^+_\omega \tr{\rho V_\omega V^*_\omega}\log\left( \frac{\gamma^+_\omega \,\tr{\rho V_\omega V^*_\omega}^2}{\gamma^-_\omega\,\tr{\rho^{1/2}V^*_\omega\rho^{1/2}V_\omega}^2}\right)\right)
\end{eqnarray}
\end{theorem}

\noindent{\it Proof.}  
Replacing $X$ by $D$ in \eqref{eq:ava} and \eqref{eq:ind} and denoting $\Vava{\omega}=\unit\otimes V_\omega$, $\Vind{\omega}=V_\omega\otimes\unit$ one obtains
\begin{eqnarray}\label{eq:phiava-phiind}
\phiava(D)&=& \sum_{\omega}\left(\gamma^-_\omega\ketbra{\Vava{\omega}
r}+\gamma^+_\omega \ketbra{\Vava{\omega}^*
r}\right)\\
\phiind(D) &= &\sum_{\omega}\left(\gamma^-_\omega \ketbra{\Vind{\omega}
r}+\gamma^+_\omega \ketbra{\Vind{\omega}^*
r}\right)
\end{eqnarray}
By Corollary \ref{cor:scalar-prod}, each vector $\Vava{\omega}r$ is orthogonal to any vector $\Vava{\omega}^*r$ and each $\Vava{\omega}r$ (respectively $\Vava{\omega}^* r$) is orthogonal to $\Vava{\omega^\prime}r$ (resp. $\Vava{\omega^\prime}^*r$) with $\omega^\prime\not=\omega$. Therefore, normalising vectors $\Vava{\omega}r,\Vava{\omega}^*r$ yield an orthonormal basis of $\h\otimes\h$. In this basis  $\phiava(D)$ turns out to be a diagonal matrix with $2\times 2$ blocks associated with each Bohr frequency $\omega$ given by
\[
\left(\begin{array}{cc}
\gamma^-_\omega \tr{\rho V_\omega^*V_\omega}&0\\
0&\gamma^+_\omega \tr{\rho V_\omega V_\omega^*}
\end{array}\right)
\]
In order to write $\phiind(D)$, compute first $\Vind{\omega}r$:
\begin{eqnarray*}
\Vind{\omega}r
& = &\frac{\tr{\rho^{1/2}V^*_\omega\rho^{1/2}V^*_\omega}}
{\tr{\rho V^*_\omega V_\omega}}\Vava{\omega}r
+\frac{\tr{\rho^{1/2}V_\omega\rho^{1/2}V^*_\omega}}
{\tr{\rho V_\omega V_\omega^*}}\Vava{\omega}^*r \\
& = & \frac{\tr{\rho^{1/2}V_\omega\rho^{1/2}V^*_\omega}}
{\tr{\rho V_\omega V_\omega^*}}\Vava{\omega}^*r,
\end{eqnarray*}
since the first term is 0.
In the same way we have
\begin{eqnarray*}
\Vind{\omega}^*r
& = &\frac{\tr{\rho^{1/2}V^*_\omega\rho^{1/2}V_\omega}}
{\tr{\rho V^*_\omega V_\omega}}\Vava{\omega}r
+\frac{\tr{\rho^{1/2}V_\omega\rho^{1/2}V_\omega}}
{\tr{\rho V_\omega V_\omega^*}}\Vava{\omega}^*r \\
& = & \frac{\tr{\rho^{1/2}V^*_\omega\rho^{1/2}V_\omega}}
{\tr{\rho V^*_\omega V_\omega}}\Vava{\omega}r,
\end{eqnarray*}
Thus, by the cyclic property of the trace, we have
\[
\Vind{\omega}r=\frac{\tr{\rho^{1/2}V^*_\omega\rho^{1/2}V_\omega}}{\tr{\rho V_\omega V^*_\omega}^{1/2}}\frac{\Vava{\omega}^*r}{\left\Vert \Vava{\omega}^*r\right\Vert},\qquad \Vind{\omega}^*r=\frac{\tr{\rho^{1/2}V^*_\omega\rho^{1/2}V_\omega}}{\tr{\rho V^*_\omega V_\omega}^{1/2}}\frac{\Vava{\omega}r}{\left\Vert \Vava{\omega}r\right\Vert}
\]

It follows that, in the above orthonormal basis of $\h\times\h$, 
obtained normalising vectors  $\Vava{\omega}r,\Vava{\omega}^*r$,  $\phiind(D)$ becomes
\begin{eqnarray*}
\phiind(D)&=&\sum_\omega\left(\gamma^+_\omega \frac{\tr{\rho^{1/2}V^*_\omega\rho^{1/2}V_\omega}^2}{\tr{\rho V^*_\omega V_\omega}}\frac{\ketbra{\Vava{\omega}r}}{\left\Vert \Vava{\omega}r\right\Vert^2}\right.\\
&+&\left . \gamma^-_\omega \frac{\tr{\rho^{1/2}V^*_\omega\rho^{1/2}V_\omega}^2}{\tr{\rho V_\omega V^*_\omega}}\frac{\ketbra{\Vava{\omega}^*r}}{\left\Vert \Vava{\omega}^*r\right\Vert^2}\right)
\end{eqnarray*}
and it turns out to be a matrix with $2\times 2$ diagonal  blocks 
associated with each Bohr frequency $\omega$ given by
\[
\left(\begin{array}{cc}
\gamma^+_\omega \frac{\tr{\rho^{1/2}V^*_\omega\rho^{1/2}V_\omega}^2}{\tr{\rho V^*_\omega V_\omega}} &0\\
0&\gamma^-_\omega \frac{\tr{\rho^{1/2}V^*_\omega\rho^{1/2}V_\omega}^2}{\tr{\rho V_\omega V^*_\omega}}\end{array}  \right)
\]
and our entropy production formula follows immediately.
\hfill $\square$

\section{Global and local equilibrium}\label{sect:global-local}

In this section we show that the entropy production 
\eqref{eq:ep-formula} vanishes if and only if the all the semigroups 
generated by each $\Ll_\omega$ satisfy the SQDB-$\Theta$ 
condition. 

It is useful to introduce some notation that allows us to focus 
more clearly contributions of each QMS generated by $\Ll_\omega$ 
to the entropy production:
\[
\nu_\omega^-=\tr{\rho V_\omega^*V_\omega},\qquad \nu^+_\omega=\tr{\rho V_\omega V^*_\omega},\qquad
\mu_\omega=\tr{\rho^{1/2}V^*_\omega\rho^{1/2}V_\omega} .
\]
In this notation the entropy production is written as
\begin{equation}
\eprod{\Tt,\rho}
=\sum_\omega\left(\gamma^-_\omega\nu^-_\omega
\log\left(\frac{\gamma^-_\omega\nu^{-\,2}_\omega}{\gamma^+_\omega\mu^2_\omega}\right)
+\gamma^+_\omega\nu^+_\omega
\log\left(\frac{\gamma^+_\omega\nu^{+\,2}_\omega}
{\gamma^-_\omega\mu^2_\omega}\right)\right)
\end{equation}
Note that, by the Schwarz inequality,
\begin{equation}\label{eq:Schwarz}
\mu^2_\omega\leq \nu^+_\omega\nu^-_\omega.
\end{equation}
Moreover
\[
\log \left( \frac{\gamma^\mp_\omega\nu^{\mp\,2}_\omega}{\gamma^\pm_\omega\mu^2_\omega} \right)=\log\left(\frac{\gamma^\mp_\omega \nu^\mp_\omega}{\gamma_\omega^\pm\nu^\pm_\omega}\right)+\log\left(\frac{\nu^+_\omega\nu^-_\omega}{\mu^2_\omega}\right)   
\]
so that we can rewrite the entropy production as 
\begin{eqnarray*}
\eprod{\Tt,\rho}&=&\sum_\omega\left(\gamma^-_\omega\nu^-_\omega-\gamma^+_\omega\nu^+_\omega\right)\log\left(\frac{\gamma^-_\omega\nu^{-}_\omega}{\gamma^+_\omega\nu^+_\omega}\right)\\
&+&\left(\gamma^-_\omega\nu^-_\omega+\gamma^+_\omega\nu^+_\omega\right)
\log\left(\frac{\nu_\omega^{+}\nu^{-}_\omega}{\mu^2_\omega}\right)
\end{eqnarray*}

\begin{corollary}\label{cor:ep=0iff}
The entropy production is zero if and only if
 $\gamma^-_\omega\nu^-_\omega=
\gamma^+_\omega\nu^+_\omega$ and 
$\nu^-_\omega\nu^+_\omega=\mu^2_\omega$.
\end{corollary}

\noindent{\it Proof.} It suffices to note that $\log({\nu_\omega^{+}\nu^{-}_\omega}/{\mu^2_\omega})$ 
is non-negative by \eqref{eq:Schwarz} and 
$(t-s)\log(t/s)$ is non-negative for all reals $t,s$ with $t\not= s$.
\hfill $\square$

\medskip

The following result shows that QMSs of stochastic 
limit type have zero entropy production of and only if the 
standard quantum detailed balance condition with reversing 
map $\Theta$ (SQDB-$\Theta$ condition) holds. 
This is not true for an arbitrary QMS as shows Example 7.3 
in \cite{FFRR-ep}.

\begin{theorem}\label{th:SQDB-ep-zero-iff}
Assume that $V_\omega$ and $H_\omega$ are real matrices for 
all $\omega$, so that the semigroup commutes with the reversing 
map $\Theta$. Then the following are equivalent:
\begin{enumerate}
\item the entropy production is zero,
\item $\rho^{1/2}L_{2\ell-1}^*= L_{2\ell} \,\rho^{1/2}$ 
for all $\ell=1,\dots, 2b$,
\item $\left(\gamma_\omega^{+}\right)^{1/2} 
\rho^{1/2}V_\omega 
= \left(\gamma_\omega^{-}\right)^{1/2} 
 V_\omega \rho^{1/2}$ for all $\omega$, 
\item the SQDB-$\Theta$ condition holds.
\end{enumerate}
\end{theorem}

{\it Proof.} $2.\Leftrightarrow 3.$ Clear from the definition 
\eqref{eq:GKSL-L} of $L_{2\ell}$ and $L_{2\ell+1}$. 
Indeed,  $ \left(\gamma_\omega^{+}\right)^{1/2} 
\rho^{1/2}V_\omega =\rho^{1/2}L_{2\ell-1}^* $  and 
$\left(\gamma_\omega^{-}\right)^{1/2}  V_\omega \rho^{1/2}
= L_{2\ell}\rho^{1/2}$ for all $\ell=1,\dots, b$. \\
$1.\Rightarrow 3.$ By Corollary \ref{cor:ep=0iff} 
we have $\nu^-_\omega\nu^+_\omega=\mu^2_\omega$ 
and so the Schwarz inequality \eqref{eq:Schwarz} turns out to be 
an equality. It follows that the operators $V_\omega\rho^{1/2}$ 
and $V_\omega\rho^{1/2}$, thought of as vectors in 
the Hilbert space of Hilbert-Schmidt operators on $\h$ are 
parallel, i.e. $\rho^{1/2}V_\omega = c_\omega 
V_\omega \rho^{1/2}$ for some constant $c_\omega$. 
Computing the scalar product with $V_\omega \rho^{1/2}$ we 
immediately find $ \mu_\omega = c_\omega \nu^-_{\omega}$, 
i.e., since $\mu^2_\omega=\nu^-_\omega\nu^+_\omega$,
$ c_\omega =\left( \nu^+_{\omega}/\nu^-_{\omega}\right)^{1/2}$
so that
\[
\rho^{1/2}V_\omega = 
\left(\frac{ \nu^+_{\omega}} { \nu^-_{\omega}}\right)^{1/2}
V_\omega \rho^{1/2}
\]
and $3.$ follows from $\gamma_\omega^-\nu_\omega^-
=\gamma_\omega^+\nu_\omega^+$. \\
$3. \Rightarrow 4.$ The SQDB-$\Theta$ condition is 
characterised by Theorem 8 in \cite{FFVU}, namely Theorem 
\ref{th:SQDB-TR} in this paper. Now, the identity $\rho^{1/2}
\theta G^*\theta = G\rho^{1/2}$ holds because we have 
assumed that $V_\omega$ and $H_\omega$ are real matrices.
Moreover, since $L_\ell=\theta L_\ell \theta$ and
$L_\ell^*=\theta L_\ell^* \theta$ for all $\ell\ge 1$, 
condition 2 of Theorem \ref{th:SQDB-TR} holds choosing 
as unitary self-adjoint the operator $u$ flipping even 
and odd indexes $\ell$ i.e. $u_{kj}= 1$ if either 
$k=2\ell$ and $j=2\ell -1$ or $k=2\ell-1$ and $j=2\ell$ 
and $u_{kj}= 0$ otherwise. \\
$4.\Rightarrow 1.$ 
For all vector $v=\sum_{\alpha,\beta} v_{\alpha\beta}\,
\theta e_\alpha\otimes e_\beta$ we have 
\begin{equation}\label{eq:vphiavav}
\left\langle v, \phiava(D) v\right\rangle 
= \sum_{\ell,j,k,\beta,\beta'} 
\overline{v}_{j\beta'}v_{k\beta} 
\left\langle e_{\beta'}, L_\ell \rho^{1/2}e_j\right\rangle 
\left\langle L_\ell\rho^{1/2}e_k , e_\beta \right\rangle 
\end{equation}
and also, by the properties of the antiunitary $\theta$
\begin{eqnarray*}
\left\langle v, \phiind(D) v\right\rangle 
& = & \sum_{\ell,j,k,\alpha,\alpha'} 
\overline{v}_{\alpha'j}v_{\alpha k} 
\left\langle \theta e_{\alpha'}, 
 L_\ell \rho^{1/2}\theta e_j\right\rangle 
\left\langle L_\ell\rho^{1/2}\theta e_k , 
\theta e_\alpha \right\rangle \\
& = & \sum_{\ell,j,k,\alpha,\alpha'} 
\overline{v}_{\alpha'j}v_{\alpha k} 
\left\langle \theta  L_\ell \theta \rho^{1/2}e_j, e_{\alpha'}\right\rangle 
\left\langle e_\alpha,  
\theta L_\ell\theta \rho^{1/2}e_k \right\rangle \\
& = & \sum_{\ell,j,k,\alpha,\alpha'} 
\overline{v}_{\alpha'j}v_{\alpha k} 
\left\langle e_j, 
\rho^{1/2}\theta  L_\ell^* \theta e_{\alpha'}\right\rangle 
\left\langle \theta L_\ell^*\theta \rho^{1/2}e_\alpha,  
e_k \right\rangle 
\end{eqnarray*}
Now, the SQDB-$\Theta$ condition holds, then 
$\rho^{1/2}\theta L_\ell^* \theta = \sum_m 
u_{\ell m} L_m \rho^{1/2}$ for a unitary self-adjoint 
$(u_{\ell m})_{1\le \ell,m \le 2b}$ so that, 
$\sum_\ell\overline{u}_{\ell m'} u_{\ell m} =\delta_{m'm}$ and
\begin{eqnarray*}
\left\langle v, \phiind(D) v\right\rangle & = & 
\sum_{\ell,j,k,\alpha,\alpha',m,m'} 
\overline{v}_{\alpha'j}v_{\alpha k} 
\overline{u}_{\ell m'} u_{\ell m} 
\left\langle e_j, 
 L_{m'} \rho^{1/2}e_{\alpha'}\right\rangle 
\left\langle L_m \rho^{1/2}e_\alpha,  
e_k \right\rangle  \\
& = &
\sum_{j,k,\alpha,\alpha',m} 
\overline{v}_{\alpha'j}v_{\alpha k} 
\left\langle e_j, 
 L_{m} \rho^{1/2}e_{\alpha'}\right\rangle 
\left\langle L_m \rho^{1/2}e_\alpha,  
e_k \right\rangle.  
\end{eqnarray*}
Changing indexes and comparing with \eqref{eq:vphiavav}, 
by the arbitrarity of $v$,  we find $\phiava(D)=\phiind(D)$ 
and the entropy production, given by \eqref{eq:ep-formula}, 
is zero.  \hfill $\square$

\medskip
{\bf Remark.} It is worth noticing here that conditions of 
Theorem \ref{th:SQDB-ep-zero-iff} are also equivalent 
to the QDB-$\Theta$ condition and so in our class of QMSs of 
stochastic limit type  the SQDB-$\Theta$ and QDB-$\Theta$. 
Indeed, since the modular group is given by $\sigma_{t}(x) 
=\rho^{\mi t} x \rho^{-\mi t}$, the  identity 
$\left(\gamma_\omega^{+}\right)^{1/2} 
\rho^{1/2}V_\omega 
= \left(\gamma_\omega^{-}\right)^{1/2} 
 V_\omega \rho^{1/2}$ 
reads $\sigma_{-\mi/2}(V_\omega) 
= \left(\gamma_\omega^-/\gamma_\omega^+\right)^{1/2}
V_\omega$.  Taking the adjoint of $\left(\gamma_\omega^{+}\right)^{1/2} 
\rho^{1/2}V_\omega 
= \left(\gamma_\omega^{-}\right)^{1/2} 
 V_\omega \rho^{1/2}$  we find also, in the same way,
$\sigma_{-\mi/2}(V_\omega^*) 
= \left(\gamma_\omega^+/\gamma_\omega^-\right)^{1/2}
V_\omega^*$.
It follows that
\begin{eqnarray*}
\sigma_{-\mi}(V_\omega) & \kern-2truept =\kern-2truept & 
\sigma_{-\mi/2}\left(\sigma_{-\mi/2}(V_\omega)\right)
=\left( \gamma_\omega^-/\gamma_\omega^+\right)^{1/2}
\sigma_{-\mi/2}(V_\omega) =
\left( \gamma_\omega^-/\gamma_\omega^+\right) V_\omega \\
\sigma_{-\mi}(V_\omega^*) & \kern-2truept =\kern-2truept & 
\sigma_{-\mi/2}\left(\sigma_{-\mi/2}(V_\omega^*)\right)
=\left( \gamma_\omega^+/\gamma_\omega^-\right)^{1/2}
\sigma_{-\mi/2}(V_\omega^*) =
\left( \gamma_\omega^+/\gamma_\omega^-\right) V_\omega^* 
\end{eqnarray*} 
and
\begin{eqnarray*}
\sigma_{-\mi}(L_{2\ell}) = 
\left( \gamma_\omega^-/\gamma_\omega^+\right) L_{2\ell}, 
& \qquad  &
\sigma_{-\mi}(L_{2\ell+1}) = 
\left( \gamma_\omega^+/\gamma_\omega^-\right) L_{2\ell-1}\\
\sigma_{-\mi}(L_{2\ell}^*) = 
\left( \gamma_\omega^+/\gamma_\omega^-\right) L_{2\ell}^*, 
& \qquad  &
\sigma_{-\mi}(L_{2\ell+1}^*) = 
\left( \gamma_\omega^-/\gamma_\omega^+\right) L_{2\ell-1}^*.
\end{eqnarray*}
Straightforward computations show that 
$\sigma_{-\mi}(H_\omega) = H_\omega$. 
It follows then from Theorem 9  in \cite{FFVU07}, that 
the QDB-$\Theta$ condition holds.

\medskip
Theorem \ref{th:SQDB-ep-zero-iff} and the above remark lead us 
to the following result essentially showing that $\rho$ is an 
equilibrium state for the QMS generated by $\Ll$ if and only if it 
is an equilibrium state for the QMSs generated by 
\emph{each} $\Ll_\omega$.

\begin{theorem}\label{th:GDB-trivial}
Let $\Ll$ be the generator of a QMS as in Section \ref{sect:QMSclass}, 
let $\rho$ be a faithful invariant state. Assume that $V_\omega$ 
is a real matrix for all $\omega$ and $H_\omega$ is a linear 
combination of $V_\omega^* V_\omega$ and $V_\omega 
V_\omega^*$. 
Then the following are equivalent: 
\begin{enumerate}
\item the QMS generated by $\Ll$ satisfies the SQDB-$\Theta$ 
condition,
\item for all $\omega$, the QMSs generated by each $\Ll_\omega$ 
admits $\rho$ as invariant state  and satisfies the SQDB-$\Theta$ 
condition.
\end{enumerate}
\end{theorem}

\noindent{\it Proof.} Clearly 2. $\Rightarrow 1.$

Conversely, if the QMS generated by $\Ll$ satisfies the SQDB-$\Theta$ 
condition, then by Theorem \ref{th:SQDB-ep-zero-iff} and the 
above Remark we have 
\begin{eqnarray*}
\rho V_\omega^* V_\omega \rho^{-1}
& = & \rho V_\omega^*\rho^{-1}\, \rho V_\omega \rho^{-1}
= \frac{\gamma_\omega^+}{\gamma_\omega^{-}}\, V_\omega^*
\,\frac{\gamma_\omega^{-}}{\gamma_\omega^{+}}\, V_\omega
= V_\omega^* V_\omega
\end{eqnarray*}
and so $\rho$ commutes with $V_\omega^*V_\omega$. In the same 
way, we can check that it commutes with $V_\omega V_\omega^*$.  
As a consequence, by the commutation rules found in the above 
Remark $\rho V_\omega^* = ({\gamma_\omega^+}/ 
{\gamma_\omega^{-}}) V_\omega^*\rho$ and 
$\rho V_\omega = ({\gamma_\omega^{-}}/ 
{\gamma_\omega^{+}}) V_\omega\rho$ and we have 
\begin{eqnarray*}
& & G_\omega \rho + \gamma_\omega^{-}V_\omega\rho V_\omega^*
+ \gamma_\omega^+V_\omega^*\rho V_\omega 
+ \rho G_\omega^* \\
&  & = (G_\omega + G_\omega^*) \rho 
+ \gamma_\omega^{+}V_\omega V_\omega^* \rho
+ \gamma_\omega^{-}V_\omega^* V_\omega \rho  \\
& & = \left( G_\omega + G_\omega^*
+\gamma_\omega^{+}V_\omega V_\omega^*
+ \gamma_\omega^{-}V_\omega^* V_\omega\right) \rho = 0.
\end{eqnarray*}
Thus $\rho$ is an invariant state for the QMS generated by 
$\Ll_\omega$.  This semigroup also satisfies the SQDB-$\Theta$ 
condition because, from $\left(\gamma_\omega^{+}\right)^{1/2} 
\rho^{1/2}V_\omega 
= \left(\gamma_\omega^{-}\right)^{1/2} 
 V_\omega \rho^{1/2}$, condition 2 of Theorem 
\ref{th:SQDB-TR} follows immediately.
\hfill $\square$ 

\medskip

\noindent{\bf Remark.} If we drop the assumptions on matrices 
$V_\omega$  and $H_\omega$ similar result holds considering 
the quantum detailed balance condition without reversing operation 
$\Theta$. In this case, however, the forward $\avanti_t$ and 
backward $\indietro_t$ states used to define the entropy 
production, defined in the same way without transpositions, 
must be thought of as states on the tensor product 
of the \emph{opposite} algebra $\mathcal{B}(\h)^{\rm o}$ 
with $\mathcal{B}(\h)$ (see \cite{FFRR-ep} Remark 2).

\section*{Acknowledgements}
Financial support from FONDECYT 1120063, ``Stochastic Analysis Network'' CONICYT-PIA grant ACT 1112 and MIUR-PRIN project
2010MXMAJR ``Evolution differential problems: deterministic 
and stochastic approaches and their interactions'' are gratefully 
acknowledged.


\begin{thebibliography}{}
%
%


\bibitem{AcFaQu}
Accardi,~L., Fagnola,~F.,  Quezada,~R.,
Weighted Detailed Balance and Local KMS Condition
for Non-Equilibrium Stationary States, 
{Bussei Kenkyu}  \textbf{97}, (2011) 318-356.


\bibitem{AcLuVo} 
L. Accardi, Y. G. Lu and I. Volovich, 
\textit{Quantum theory and its stochastic 
limit}, Springer-Verlag, Berlin, (2002).

\bibitem{Agarwal}
 Agarwal,~G.S., Open quantum Markovian systems and the microreversibility. {\it Z. Physik}  \textbf{258} (1973) 409--422.


\bibitem{Alicki-Lendi}
Alicki,~R., Lendi,~K., \textit{Quantum Dynamical Semigroups 
and Applications}, \textit{Lecture Notes in Physics} \textbf{286}, 
Springer-Verlag, Berlin 1987.

\bibitem{BolQue}
Bola\~nos, J., Quezada, R., A cycle decomposition and entropy 
production for circulant quantum Markov semigroups. 
{\it Infin. Dimens. Anal. Quantum Probab. Relat. Top.}
{\bf 16} 1350016 (2013).

\bibitem{Breuer}
Breuer, H.P., 
\newblock Quantum jumps and entropy production.
\newblock {\it Phys. Rev. A}, {\bf 68} (2003), 032105.

\bibitem{MR2046709}
Callens, I., De~Roeck, W., Jacobs, T.,  Maes, C., 
Neto{\v{c}}n{\'y}, K.,
\newblock Quantum entropy production as a measure 
of irreversibility.
\newblock {\it Phys. D}, \textbf{187} (2004) 383--391.

\bibitem{CaFaHa}
Carbone, R., Fagnola, F., Hachicha, S., 
Generic quantum Markov semigroups: 
the Gaussian gauge invariant case. 
{\it Open Syst. Inf. Dyn.} {\bf 14} (2007), 425-444.

\bibitem{Cipriani}
Cipriani, F., Dirichlet forms and markovian semigroups on 
standard forms of von Neumann algebras.
{\it J. Funct. Anal.} \textbf{147},  (1997) 259--300.

\bibitem{Qian2003}
Da-Quan Jiang, Min Qian, and Fu-Xi Zhang.
\newblock Entropy production fluctuations of finite 
{M}arkov chains.
\newblock {\it J. Math. Phys.} {\bf 44}, (2003) 4176--4188.

\bibitem{Dav}
Davies, E.B., Markovian master equations. 
{\it Commun. Math. Phys.} {\bf 39} (1974) 91--110.

\bibitem{DeFr}
Derezynski, J., Fruboes, R., Fermi golden rule and open quantum 
systems, in: {\it Open Quantum Systems III - 
Recent Developments}, 
Lecture Notes in Mathematics {\bf 1882}, 
Springer Berlin, Heidelberg (2006), pp. 67–116.

\bibitem{DuSn}
Duvenhage, R., Snyman, M., Detailed balance and entanglement 
{\it J. Phys. A: Math. Theor. } {\bf 48} (2015) 155303. 

\bibitem{FFRR-LNM1882}
Fagnola, F., Rebolledo, R., Notes on the Qualitative Behaviour 
of Quantum Markov Semigroups, in:  \textit{Open Quantum 
Systems III - Recent Developments}. Lecture Notes in 
Mathematics \textbf{1882}, Springer Berlin, Heidelberg (2006), 
pp. 161--206.

\bibitem{FFRR-QP29}
Fagnola, F., Rebolledo, R., {From classical to quantum 
entropy production}, in  \textit{Quantum Probability and 
Infinite Dimensional Analysis},  QP-PQ: Quantum Probability 
and White Noise Analysis \textbf{25},  
 World Scientific, Singapore (2010), pp. 245--261.

\bibitem{FFRR-ep}
Fagnola, F., Rebolledo, R., Entropy production for quantum 
Markov semigroups. 
{\it Commun. Math. Phys.} {\bf 335}, 547--570 (2015). 
doi 10.1007/s00220-015-2320-1

\bibitem{FFVU07}
Fagnola, F., Umanit\`a, V.: Generators of detailed balance 
quantum Markov semigroups. {\it Inf. Dim. Anal. Quant. Probab. 
Relat. Top.} {\bf 10},  (2007) 335--363.

\bibitem{FFVU08}
F.~Fagnola, V.~Umanit\`a, Detailed Balance, time reversal 
and generators of Quantum Markov Semigroups, {\it M. Zametki}, 
{\bf 84} (1), 108-116 (2008) (in Russian). English translation  
{\it Math. Notes}  {\bf 84} (1-2) 108-115 (2008).

\bibitem{FFVU}
Fagnola, F., Umanit\`a, V., 
Generators of KMS Symmetric Markov Semigroups on $\bo{\h}$
Symmetry and Quantum Detailed Balance. 
{\it Commun. Math. Phys.} {\bf 298}, (2010) 523--547. 

\bibitem{FFVU12}
Fagnola, F., Umanit\`a, V., 
Generic Quantum Markov Semigroups, Cycle Decomposition 
and Deviation From Equilibrium, {\it Infin. Dimens. Anal. 
Quantum Probab. Relat. Top.}, {\bf 15}
No. 3 (2012) 1250016 (17 pages).


\bibitem{JP-2001}
Jak{\v s}i{\'c}, V., Pillet, C.-A.,
On entropy production in quantum statistical mechanics.
{\it Commun. Math. Phys.} \textbf{217} (2001), (2001) 285--293.  

\bibitem{JP-2013}
Jak{\v s}i{\'c}, V., Pillet, C.-A. and Westrich, M., 
Entropic fluctuations of quantum dynamical semigroups. 
{\tt arxiv arXiv:1305.4409}


\bibitem{KFGV}
Kossakowski, A., Frigerio, A., Gorini V., Verri, M.,
Quantum detailed balance and KMS condition.
{\it  Comm. Math. Phys.}  {\bf 57},  (1977)  97--110.

\bibitem{MRM}
Maes, C., Redig, F., Van~Moffaert, A.,
\newblock On the definition of entropy production, via examples.
\newblock {\em J. Math. Phys.}, {\bf 41} (2000), 1528--1554.  

\bibitem{Maje}
Majewski,~W.A., 
The detailed balance condition in quantum statistical mechanics,
{\it J. Math. Phys.} \textbf{25}, (1984) 614--616. 

\bibitem{MaSt}
Majewski,~W.A., Streater,~R.F., Detailed balance and quantum dynamical maps,  {\it J. Phys. A: Math. Gen.} \textbf{31},  (1998) 
7981--7995.

\bibitem{Onsager:1931p1766}
Onsager, L., Reciprocal relations in irreversible processes. I.
{\it Phys Rev} {\bf 37}, (1931) 405--426.

\bibitem{Partha}  
Parthasarathy,~K.R.,  
\textit{An introduction to quantum stochastic calculus}, 
{Monographs in Mathematics} \textbf{85}, 
Birkh\"auser-Verlag, Basel 1992.



\bibitem{Seif}  
Seifert, U., Stochastic thermodynamics, fluctuation theorems,
and molecular machines, {\it Rep. Progr. Phys.} {\bf 75} (2012) 
126001 



\end{thebibliography}
\end{document}